\documentclass{ws-mplb1}

\begin{document}

\markboth{Lakhno V.D.}
{Cooper pairs and bipolarons}

%%%%%%%%%%%%%%%%%%%%% Publisher's Area please ignore %%%%%%%%%%%%%%%
%
\catchline{}{}{}{}{}
%
%%%%%%%%%%%%%%%%%%%%%%%%%%%%%%%%%%%%%%%%%%%%%%%%%%%%%%%%%%%%%%%%%%%%

\title{COOPER PAIRS AND BIPOLARONS}

\author{VICTOR LAKHNO}

\address{Institute of Mathematical Problems of Biology RAS \\
the Branch of Keldysh Institute of Applied Mathematics of Russian Academy of Sciences\\
142290, Vitkevicha str. 1,  Pushchino, Moscow Region, Russia\\
lak@impb.psn.ru}

\maketitle

%\begin{history}
%\received{Day Month Year}
%\revised{Day Month Year}
%\accepted{(Day Month Year)}
%\comby{(xxxxxxxxxx)}
%\end{history}

\begin{abstract}
It is shown that Cooper pairs are a solution of the bipolaron problem for model Fr\"{o}hlich Hamiltonian. The total energy of a pair for the initial Fr\"{o}hlich Hamiltonian is found. Differences between the solutions for the model and initial two-particle problems are discussed
\end{abstract}


\section {Introduction}

The theory of Cooper pairs\cite{lit1} appeared earlier than the superconductivity theory did.\cite{lit2}
It forms the basis for treating the superconductivity occurrence process at the microscopic level. Central to the theory is the mechanism of electron-phonon interaction (EPI) which provides attraction between a pair of electrons. The same mechanism is fundamental for the theory of bipolarons, i.e. two electrons bounded by electron-phonon interaction.

Since a Cooper pair is considered to involve a lot of centers of other pairs, it has never been treated as a bipolaron. This fact has dramatically affected the way of developing the superconductivity theory of ordinary superconductors described by BCS theory.

This situation is just analyzed in the paper.

\section {Cooper consideration}

The problem of two electrons in a phonon field is originally solved in Cooper pioneering paper.\cite{lit1} The fact that the system involves many particles does not affect the validity of the two-particle approximation since, in view of Pauli principle, electrons under the Fermi surface only slightly disturb the state of electrons occurring outside the Fermi surface. The aim of Cooper paper was to show that the two-particle problem has a spectrum in which the ground state is separated from the quasicontinuous spectrum by a gap and occurs below the Fermi level.

 To solve this problem Cooper made an important simplifying assumption – instead of initial EPI he used a truncated Hamiltonian where nonzero interaction exists only for the wave vectors of electrons occurring in a narrow layer of energies near the Fermi surface. Besides, in the model Hamiltonian, Coulomb repulsion between paired electrons was not taken into account since it was thought to be screened by rearrangement of the electron gas. The use of the truncated Hamiltonian automatically leads to the appearance of an isolated energy level which occurs inside the Fermi surface in the case of attracting EPI.  The question of the energy advantage of this level, i.e. the question of the total energy of the two-particle system in a phonon field was not considered at all.

An answer to this question was given already in the many-particle BCS theory, where with the use of the Cooper model Hamiltonian a many-electron wave function composed of paired wave functions was constructed which yielded a lower energy value than the wave function without paired states did. The fact that the paired wave functions involved in the BCS wave function had no concern with the paired wave functions of the Cooper problem was of no importance. More likely this caused a problem in writing numerous manuals on microscopic superconductivity theory which traditionally start with a presentation of the Cooper two-particle problem.

Discovery of high-temperature superconductors demonstrated that the BCS theory is not applicable in their case. This, in turn, requires a more detailed analysis of the concept of Cooper pairs. Discontent with the Cooper theory gave birth to many alternative theories of pairing. At present, however, there is no cause for discarding the superconductivity mechanism on the basis of EPI.

\section {Bipolaron theory of Cooper pairing}

Thus, the initial Hamiltonian in the Cooper problem is Fr\"{o}hlich Hamiltonian which in the coordinates of the center of mass has the form:

\begin{eqnarray}\label{1}
\hat{H}=-\frac{\hbar^2}{2M_e}\Delta _R-\frac{\hbar^2}{2\mu_e}\Delta _r+U(r)+
\sum\hbar\omega_ka^+_ka_k+\sum_k2\cos\frac{\vec{k}\vec{r}}{2}\left(V_ke^{i\vec{k}\vec{R}}a_k+H.c.\right),
\end{eqnarray}
where $R$, $r$ are coordinates of the center of mass and relative coordinates of electrons, respectively;
$M_e=2m$,  $\mu_e=m/2$, $m$ is the electron effective mass;
$a^+_k$, $a_k$ are operators of the phonon field; for a polaron medium
$V_k=(e/k)\sqrt{2\pi\hbar\omega/\tilde{\epsilon}V}$, $\tilde{\epsilon}^{-1}=\epsilon^{-1}_{\infty}-\epsilon^{_1}_0$,
$\omega_k=\omega$ is the phonon frequency, $e$ is the electron charge; $\epsilon_{\infty}$, $\epsilon_0$ are high-frequency and static dielectric constants, $V$ is the system's volume; $U(r)$ is the Coulomb interaction between electrons. Notice that Cooper dealt with acoustical phonons, which are actual for ordinary superconductors. For high-temperature superconductors, the use of Fr\"{o}hlich Hamiltonian for optical phonons is more suitable.  For Cooper, however, in view of his model approximation, this fact was immaterial.

After elimination of the coordinates of the center of mass $\vec{R}$  via Heisenberg transformation, with the use of Lee-Low-Pines transformation,\cite{lit3}
\begin{eqnarray}\label{2}
\hat{S}=\exp\left\{\sum_k f_k\left(a_k-a^+_k\right)\right\}
\end{eqnarray}
the energy of electron-phonon interaction, according to (1) takes the form:
\begin{eqnarray}\label{3}
U_{int}(r)=\left\langle 0\left|S^{-1}\left(2\sum_kV_k\cos\frac{\vec{k}\vec{r}}{2}\left(a_k+a^+_k\right)\right)S\right|0\right\rangle=
4\sum_kV_kf_k\cos\frac{\vec{k}\vec{r}}{2}.
\end{eqnarray}
Let us find an explicit form of $U_{int}(r)$  in the limit of weak EPI, i.e. in the case which was considered by Cooper\cite{lit1} (in the limit of strong EPI the form of $U_{int}(r)$  is given there.\cite{lit5}) Since in this case the problem is solved within the perturbation theory where for a zero approximation, $f_k$  corresponding to weak-coupling polaron states are chosen:\cite{lit3}
\begin{eqnarray}\label{4}
f_k=-\frac{V_k}{\hbar\omega_k+\hbar^2k^2/2m},
\end{eqnarray}
then with the use of (3) and (4) we express $U_{int}(r)$  in the form:
\begin{eqnarray}\label{5}
U_{int}(r)=-\frac{4e^2}{\tilde{\epsilon}r}\left(1-e^{-r/2r_0}\right),
\end{eqnarray}
where
\begin{eqnarray}\label{6}
r_0=\left(\hbar/2m\omega\right)^{1/2},
\end{eqnarray}
$r_0$  has the meaning of a characteristic size in the polaron theory.

Expression (5) yields a straightforward conclusion made by Cooper: interaction between the electrons is attracting and Schr\"{o}dinger equation corresponding to potential (5) always has a discrete level lying below the Fermi surface. The latter follows from the fact that for $r\rightarrow\infty$  the electron interaction potential has a Coulomb form which automatically provides the existence of a discrete level with negative energy.

As was noted above, Cooper did not take into account Coulomb repulsion of electrons. If it is taken into consideration, the total interaction potential $U_{tot}$  takes the form:

\begin{eqnarray}\label{7}
U_{tot}(r)=U_{int}(r)+U(r),
\end{eqnarray}
In the absence of screening $U(r)=e^2/\epsilon_{\infty}r$  the discrete level exists on condition that
$3\epsilon_0>4\epsilon_{\infty}$. In the general case for $U(r)$ one should use the screening expression. For example, in Thomas-Fermi approximation $U(r)=(e^2/\epsilon_{\infty}r)\exp (-r/r_{TF})$, where $r_{TF}$  is Thomas-Fermi radius. This changes the condition of the existence of a discrete level making it less rigid. Notice, that Cooper discarded expression (5) and used instead a simplified expression for Fourier components of the interaction potential $U_{int}(k)=v/V$  for  $E_F\leq\hbar^2k^2/2m\leq E_F+\delta$,  $v=const$ and $U_{int}(k)=0$  for other values of $k$ which leads to interaction  $U_{int}(r)\sim \sin\left(\sqrt{2\mu_e E_F}r/\hbar\right)/r$  and the energy of the discrete level $\Delta$ :
\begin{eqnarray}\label{8}
\Delta=\delta \exp \left[-1/v\rho\left(E_F\right)\right],
\end{eqnarray}
which corresponds to the state radius $\bar{r}\approx\hbar^2k_F/m\Delta$, where $\rho _f(E_F)$ is the density of the states at the Fermi level.

To answer the question of the value of the total energy in the Cooper problem let us consider the expression for the bipolaron total energy in the weak coupling approximation:\cite{lit4}
\begin{eqnarray}\label{9}
E=\Delta E+2\sum_k\bar{V}_kf_k+\sum f^2_k+\bar{T}+\bar{U},
\end{eqnarray}
$$\bar{T}=\left\langle \Psi\left|-\frac{\hbar^2}{2\mu_e}\Delta_r\right|\Psi\right\rangle,\ \ \
\bar{U}=\left\langle \Psi\left|U(r)\right|\Psi\right\rangle,$$
$$\bar{V}_k=2V_k\left\langle \Psi\left|\cos\frac{\vec{k}\vec{r}}{2}\right|\Psi\right\rangle ,$$
where $\Psi$ and $f_k$ are found from the condition of the minimum of the bipolaron energy $E$ with respect to $\Psi$ and $f_k$ . With regard to the fact that the value of recoil energy $\Delta E$ involved in (9) in the limit of weak coupling is equal to: $\Delta E=\sum\left(\hbar^2k^2 /2M_e \right)f^2_k$,\cite{lit4} we express  $f_k$ in the form:
\begin{eqnarray}\label{10}
f_k=-\frac{\bar{V}_k}{\omega_k+\hbar^2k^2/2M_e}.
\end{eqnarray}
We will seek for the minimum of (9), choosing the probe wave function $\Psi$ in the Gaussian form:
\begin{eqnarray}\label{11}
\left|\Psi (r)\right|^2=\left(\frac{2}{\pi l^2}\right)^{3/2}e^{-2r^2/l^2},
\end{eqnarray}
where $l$ is a variational paremeter. Substituting (10), (11) into (9) and minimizing the expresion obtained with respect to $l$, we express $E$ as:
\begin{eqnarray}\label{12}
E=-\frac{1}{24}\left[\frac{16}{\sqrt{\pi}}-\frac{8}{\sqrt{2\pi}}\frac{1}{(1-\eta)}\right]^2\alpha^2\hbar\omega ,
\end{eqnarray}
\begin{eqnarray}\label{13}
l=12\left(\hbar^2\tilde{\epsilon}/me^2\right)/\alpha\left[\frac{16}{\sqrt{\pi}}-\frac{8}{\sqrt{2\pi}}\frac{1}{(1-\eta)}\right],
\end{eqnarray}
$$\eta=\epsilon_{\infty}/\epsilon_0,\ \ \alpha=(e^2/\hbar\tilde{\epsilon})\sqrt{m/2\hbar\omega},$$
where $\alpha$ is a constant of EPI, $l$  has the meaning of the characteristic size of a Cooper pair.  From (12), (13) follows a condition of the existence of a discrete level (i.e. existence of a bipolaron state) in the limit
$\alpha\rightarrow0$ : $\epsilon_0>1,4\epsilon_{\infty}$  which is close to the earlier obtained criterion.

Expression (12), though corresponding to a gain in the total energy of a Cooper pair (i.e. bipolaron state), for $\alpha<1,4$ corresponds to a metastable state. The reason is that the bipolaron state (12) is not stable with respect to its decay into two individual polaron states with energy $E=-2\alpha\hbar\omega$ which is always fulfilled in the limit  $\alpha\rightarrow 0$.

Notice that expression (12) obtained in the limit of weak coupling differs from the bipolaron energy expression in the limit of strong coupling only in the numerical coefficient.

\section {Cooper pairs and BCS}

According to Cooper, only the electrons in the thin energy layer $\delta$ near the Fermi surface are paired. Hence, by Cooper, only a small portion of electrons $n'\cong(\delta/E_F)n$   in metal are paired, where $n$ is a concentration of electrons in metal.

On the contrary, the BCS theory suggests that at zero temperature all n electrons should be paired. This paradox has not been clearly explained as yet.

An answer to this question can be obtained if we consider the initial many-particle problem. To this end Fr\"{o}hlich Hamiltonian should involve interaction of all electrons with the phonon field and Coulomb repulsion of all electrons with one another. No transformation can rearrange such a Hamiltonian into a set of "effective" electron pairs. In Bogolubov theory\cite{lit6} such an "effective" Hamiltonian is just postulated. This follows from the fact that Hamiltonian of an individual pair (1) will not commutate with the total many-particle Hamiltonian. This, in turn, means that the state of an individual pair is not a motion integral. Hence, resolution of the paradox lies in the fact that we have a fluctuating picture of electron pairs (bipolarons) in a superconductor. In this picture, the portion of electrons occurring in the paired state is usually small, since the life-time of the pairs is short.

The fact that the BCS theory paints quite a different picture – there all the electrons are described by a stationary wave function composed of paired wave functions – indicates only that the choice of even a rough probe wave function yields a gain in the total energy of the system which is surely a remarkable result. This approach, however, turned out to be inapplicable in the case of high-temperature superconductors.

One of the reasons for writing this paper is the opinion of an expert community who express distrust to description of superconductivity on the basis of the bipolaron concept. From the above discussion it appears that if this is the case, the distrust is, in essence, expressed to the Cooper idea of pairing, since the latter is nothing but a very rough solution of the bipolaron problem.

Presently the polaron theory of superconductivity in its most consistent with the experiment form is presented there.\cite{lit7}

\section*{Acknowledgements}
The work was supported by RFBR, N 16-07-00305 and RSF, N 16-11-10163.

\end{document}